\documentclass[prb,preprint]{revtex4}

\usepackage{graphicx}
\begin{document}

\title{Nonreciprocal magnon propagation in a noncentrosymmetric ferromagnet}

\author{Y. Iguchi, S. Uemura, K. Ueno, and Y. Onose} 
\affiliation{Department of Basic Science, University of Tokyo, Tokyo, 153-8902, Japan}

%\date{}

\begin{abstract}
\bf{Relativistic spin-orbit interaction drastically modifies electronic band and endows emergent functionalities.   
One of the example is the Rashba effect\cite{rashba,rashba2}. In noncentrosymmetric systems such as interface\cite{nitta} and polar materials\cite{casella,ishizaka}, the electronic band is spin-splitted depending on the momentum direction owing to the spin-orbit interaction, which is useful for the electric manipulation of spin current. Similar relativistic band-modification is also emergent for spin wave (magnon) in magnetic materials. The asymmetric magnon band dispersion induced by the Dzyaloshinskii-Moriya interaction\cite{dzyaloshinskii, moriya}, which is antisymmetric exchange interaction originating from the spin-orbit interaction, is theoretically expected\cite{melcher,kataoka}, and experimentally observed recently in noncentrosymmetric ferromagnets\cite{zakeri,di}. Here, we demonstrate that the nonreciprocal microwave response can be induced by the asymmetric magnon band in a noncentrosymmetric ferrimagnet LiFe$_5$O$_8$. This result may pave a new path to designing magnonic device based on the relativistic band engineering.
}
\end{abstract}
%\pacs{}
\maketitle
Dzyaloshinskii-Moriya interaction $H_{\rm DM}={\bf D} \cdot ({\bf S}_i \times {\bf S}_j)$ originates from the relativistic spin-orbit interaction as well as exchange interaction\cite{dzyaloshinskii, moriya}.
The vector ${\bf D}$ becomes nonzero when the inversion symmetry at the midpoint of two magnetic moments is broken, and the direction is determined by the symmetry rule termed Moriya-rule\cite{moriya}. The Dzyaloshinskii-Moriya interaction modulates magnetic structures, and, in some cases, produces novel magnetic state such as skyrmion crystal\cite{muhlbauer,yu}. 
%have been frequently observed in low-symmetrical magnetic materials. For example, the helical spin structure caused by the combination of the symmetric exchange interaction and antisymmetric Dzyaloshinskii-Moriya interaction are frequently observed in the noncentrosymmetric magnets. Recently, it has been observed that the topological magnetic state of skyrmion crystal is emergent as a result of superposition of three helical spin structures in chiral helimagnets. 
Dynamical magnetic states are more largely affected by the Dzyaloshinskii-Moriya interaction. Even in the completely spin-polarized ferromagnetic state, the spin wave excitation (magnon) is affected by the Dzyaloshinskii-Moriya interaction; it acquire an additional phase factor owing to the Dzyaloshinskii-Moriya interaction in the course of the propagation, which can be viewed as the Berry phase. Recently, the Hall effect of magnons caused by the Berry phase due to the Dzyaloshinskii-Moriya interaction is observed in terms of thermal transport\cite{onose}. 
In noncentrosymmetric ferromagnets, there is the uniform component of Dzyaloshinskii-Moriya interaction and the magnons acquire the phase factor proportional to the propagation distance. As a result, 
the asymmetric magnon band is realized in this class of materials. In this paper, we report unique microwave response owing to the asymmetric magnon band in a noncentrosymmetric ferromagnet LiFe$_5$O$_8$.

\begin{figure}
\begin{center}
\includegraphics*[width=13cm]{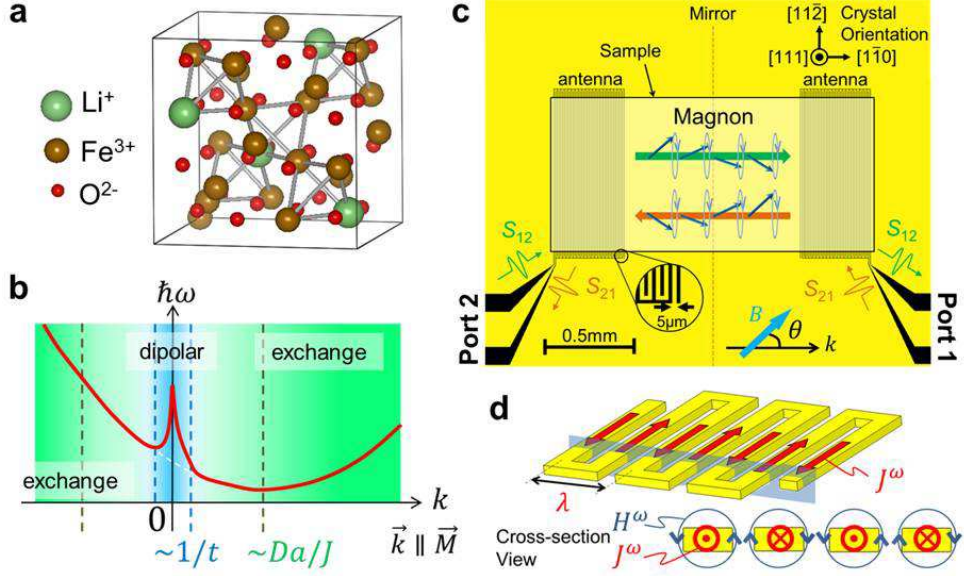}
\caption{\textbf{Experimental setup for the detection of nonreciprocal magnon propagation in noncentrosymmetric LiFe$_5$O$_8$.}
\textbf{a}, Noncentrosymmetric crystal structure of LiFe$_5$O$_8$. \textbf{b}, Sketch of the expected magnon dispersion in the ferromagnetic state of LiFe$_5$O$_8$. \textbf{c}, Detail of experimental setup. Meander microwave antennae are connected to two ports of network analyzer. The rectangular sample is put on the top of the two antennae. \textbf{d}, Illustration of the spatially oscillating magnetic fields induced by the electric current in the meander antenna. }
\end{center}
\end{figure}

LiFe$_5$O$_8$ is crystallized into a spinel-type crystal structure as shown in Fig. 1a. This material is similar to magnetite but Li ions are substituted for the one-fourth of Fe ions at the B site of spinel structure, and ordered so that the mirror symmetry is broken (space group $P4_{1}32$) \cite{braun}. The Fe magnetic moments show collinear ferromagnetic (more precisely, ferrimagnetic) order around 900 K\cite{landolt}. 
The effective Hamiltonian in this class of noncetrosymmetric material was deduced based on the symmetry analysis as followings\cite{kataoka};    
\begin{eqnarray}
H=\int d{\bf r} \Bigl[\frac{Ja^2}{2}(\nabla {\bf M})^2+Da{\bf M}\cdot (\nabla \times {\bf M})\Bigr]+H_A+H_D+H_Z.
\end{eqnarray}
Here, ${\bf M}$ is the spatially dependent magnetization, $J$ is the ferromagnetic exchange interaction, $D$ is the Dzyaloshinskii-Moriya interaction, $a$ is lattice constant, and $H_A$, $H_D$, and $H_Z$ are the magnetic anisotropy, the magnetic dipole-dipole interaction, and Zeeman energy, respectively. If $H_A$ is absent, helical spin structure is realized owing to the Dzyaloshinskii-Moriya interaction at zero magnetic field as observed in MnSi\cite{ishikawa}. Nevertheless, it seems that the collinear ferromagnetic state is stable in this system even at zero magnetic field probably because the Dzyaloshinskii-Moriya interaction is weaker than $H_A$.
The magnetic anisotropy is also not so large. The magnetization is saturated in a small magnetic field ($\sim$ 0.2 T) with any direction (see Supplementary Information).
The effect of the weak Dzyaloshinskii-Moriya interaction can be sensitively observed in terms of magnetic excitation. 
   
Figure 1b shows the expected magnon dispersion relation along the magnetization direction in the ferromagnetic state. As Kataoka theoretically suggested\cite{kataoka}, the ferromagnetic exchange and Dzyaloshinskii-Moriya interactions gives rise to asymmetric energy dispersion in ferromagnets with the space group of $P4_{1}32$. It is parabolic but the minimum of wave number $k$ is shifted by $\sim D/Ja$ from the $\Gamma$ point along the magnetization direction.
%While the static spin structure is not modified by the Dzyaloshinskii-Moriya interaction originating from the mirror-symmetry-broken chiral crystal structure probably because of the relatively large magnetic anisotropy, the magnon excitation is expected to be largely influenced by the Dzyaloshinskii-Moriya interaction. that in the ferromagnetic statesKataoka theoretically suggested  in chiral and cubic crystals,
%the magnon excitation shows the parabolic energy dispersion and the minimum wave number $k$ is nonzero ($\sim D/Ja$) along the magnetization direction owing to the Dzyaloshinskii-Moriya interaction ($D,J$, and $a$ are the magnitude of Dzyaloshinsky-moriya and ferromagentic exchange interactions, and lattice constant, respectively). 
Therefore, the nonreciprocal magnon propagation (magnons with $+k$ and $-k$ are not degenerate) is expected in this system. However, it should be noted that the classical dipole-dipole interaction $H_D$ dominates magnon dispersion around $k=0$. The magnon propagating via $H_D$ is denoted as magnetostatic wave\cite{stancil}.
For a plate-like sample, the magnetostatic waves propagating along the magnetization within the sample plane has the negative group velocity and symmetric dispersion relation. The crossover of dominant interaction from $H_D$ to exchange interactions occurs around $k \sim 1/t$, where $t$ is sample thickness. Therefore, in order to observe the nonreciprocal response of magnons, the microwave excitation of the magnons with large $k$ is needed.  

In order to observe the nonreciprocal magnon propagation, we design the special experimental setup as shown in Fig. 1c. The two meander antennae made of Au 200nm /Ti 50nm thin film are fabricated on a sapphire substrate with use of photolithography technique and electron beam evaporation.
The conducting path goes upward and turns back repeatedly 20 times in the antenna. The width and space of the conducting line are both 5 $\mu$m. As illustrated in Fig. 1d, the spatially oscillating magnetic field along the crossing direction with the period of 20 $\mu$m is induced by the electric current along the conduction line. We put the plate-like rectangular samples (LiFe$_5$O$_8$ sample and reference Y$_3$Fe$_5$O$_{12}$ sample) on the top of the two antennae. For both the samples, the widest surface is (111) plane and the longest direction is [$1\bar{1}0$].
The sample dimensions of the samples are both $2 \times 0.9 \times 0.6$ mm$^3$.  The antennae can excite and detect the magnons with the wave length $\lambda$ of 20 $\mu$m, which is much larger than the sample thickness, and are within the exchange regime shown in the Fig. 1b. 
Therefore, the effect of band asymmetry induced by the Dzyaloshinskii-Moriya interaction can be sensitively observed with this experimental setup.

\begin{figure}
\begin{center}
\includegraphics*[width=12cm]{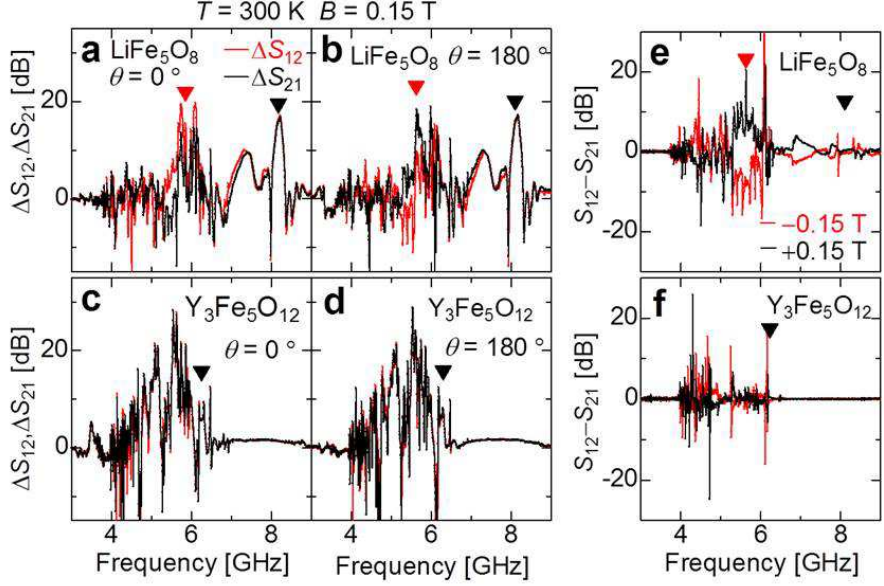}
\caption{\textbf{Microwave transmittance and its nonreciprocity} \textbf{a-d},\ Microwave transmittances owing to magnon propagation from port 2 to port 1 ($\Delta S_{12}$) and from port 1 to port 2 ($\Delta S_{21}$) for (a,b) LiFe$_5$O$_8$ and (c,d) Y$_3$Fe$_5$O$_{12}$ are plotted as a function of microwave frequency at room temperature. Magnitude of the magnetic field \textit{B} is 0.15 T, and \textit{B} is (a,c) parallel ($\theta$ =0 deg.) or (b,d) anti-parallel ($\theta$ = 180 deg.) to the direction of magnon propagation. FMR ($k \sim 0$) signals are represented by inverted black triangles and the large wave vector magnon modes are indicated by inverted red triangles. \textbf{e,f},\ Nonreciprocity of microwave transmission $S_{12} - S_{21}$ of (e) LiFe$_5$O$_8$ and (f) Y$_3$Fe$_5$O$_{12}$ at ${\bf B}=\pm0.15$ T and $\theta$ =0$^\circ$.}
\end{center}
\end{figure}

In Figs. 2a-d, we show the microwave transmittance owing to magnon propagation $\Delta S_{12}$ and $\Delta S_{21}$ for the LiFe$_5$O$_8$ sample and the reference sample Y$_3$Fe$_5$O$_{12}$ at room temperature. 
(Here, $\Delta S_{12}$ and $\Delta S_{21}$, respectively, stand for the difference of the transmission coefficient $S_{12}$ (from port 2 to port 1) and $S_{21}$ (from port 1 to port 2) from the back ground for the microwave circuit shown in Fig. 1c; for details, see Supplementary Information.)
Figures 2a,c show the data measured in a magnetic field parallel to the magnon propagation direction ($\theta=0^\circ$), and Figs. 2b,d show the data for the opposite magnetic field direction ($\theta=180^\circ$). For LiFe$_5$O$_8$, $\Delta S_{12}$ spectra is composed of several peaks. By the comparison with the result of conventional microwave absorption experiment(see Supplementary Information), the broad peak around 8.3 GHz is ascribed to $k \sim 0$ FMR mode. The lower frequency modes 
have larger $k$ because magnon around $k=0$ has negative slope of band dispersion or group velocity. Importantly, the large nonreciprocity (difference of $\Delta S_{12}$ and $\Delta S_{21}$) is observed around 5.5 GHz while the FMR mode is reciprocal. For centrosymmetric Y$_3$Fe$_5$O$_{12}$, the transmittance spectra is reciprocal in all the frequency region. When the magnetic field is reversed, the nonreciprocity is also reversed as shown in Fig. 2b. Figures 2e,f show the magnitude of nonreciprocity $S_{12}-S_{21}$ as a function of frequency for (e) LiFe$_5$O$_8$  and (f) Y$_3$Fe$_5$O$_{12}$. The large nonreciprocity is observed around 5.5 GHz for LiFe$_5$O$_8$ while the nonreciprocity is negligible in Y$_3$Fe$_5$O$_{12}$. It should be noted that there is the mirror symmetry at the center of this experimental systems as indicated by the red dashed line in Fig. 1 c, which ensures $S_{12}=S_{21}$ for Y$_3$Fe$_5$O$_{12}$ at $\theta=0$. The symmetry is broken when the LiFe$_5$O$_8$ sample is placed because of the crystal symmetry. 

\begin{figure}
\begin{center}
\includegraphics*[height=8cm]{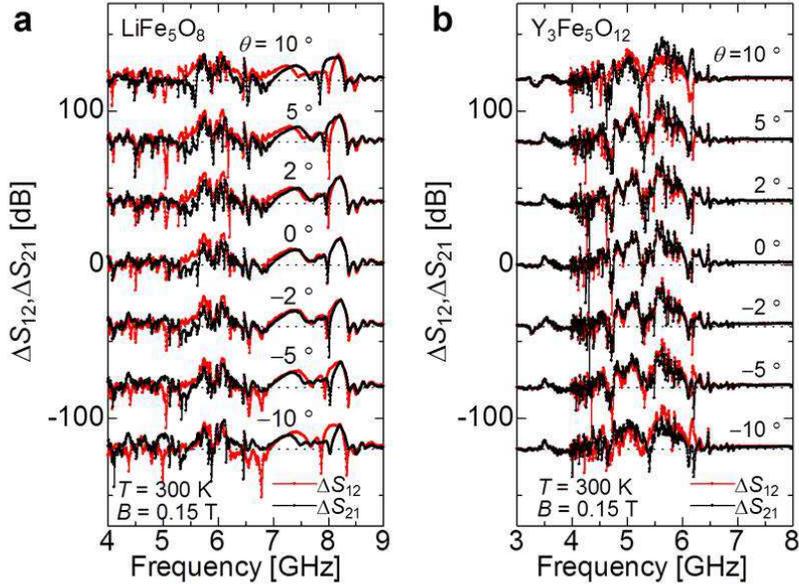}
\caption{\textbf{Angular dependence of microwave transmittance}  \textbf{a,b,} Angular dependence of $\Delta S_{12}$ and $\Delta S_{21}$ for (\textbf{a}) LiFe$_5$O$_8$ and (\textbf{b}) Y$_3$Fe$_5$O$_{12}$ at $B=0.15$ T.}
\end{center}
\end{figure}

In order to examine the microscopic origin of nonreciprocity, we plot in Figs. 3 a,b the $\Delta S_{12}$ and $\Delta S_{21}$ spectra at various magnetic field angles around $\theta=0$ for LiFe$_5$O$_8$(Fig. 3a) and Y$_3$Fe$_5$O$_{12}$ (Fig. 3b). The nonreciprocity in Y$_3$Fe$_5$O$_{12}$ is quite sensitive to the magnetic field angle. The nonreciprocal response emerges as the magnetic field is tilted from $\theta =0$, and the nonreciprocity is reversed when the tilting direction is reversed. A similar field-angle dependence is also observed for the $k=0$ mode of LiFe$_5$O$_8$ around 8.3 GHz. On the other hand, the nonreciprocity of the large $k$ mode for LiFe$_5$O$_8$ (around 5.5 GHz) is rather insensitive to the magnetic field angle. 

\begin{figure}
\begin{center}
\includegraphics*[width=8cm]{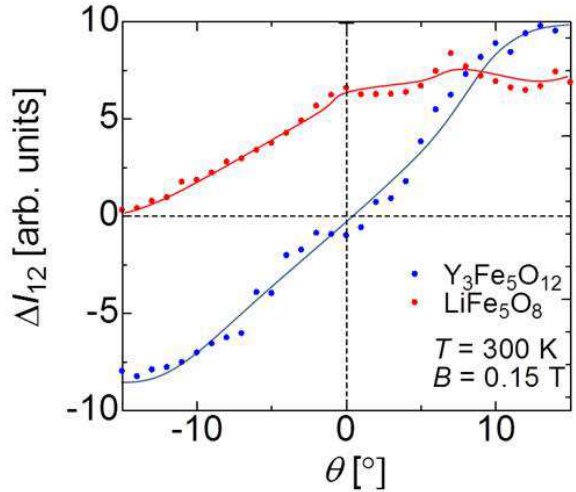}
\caption{\textbf{Integrated intensity of nonreciprocity} Integrated intensities of nonreciprocity $\Delta I_{12}$ for LiFe$_5$O$_8$ (blue circle) and Y$_3$Fe$_5$O$_{12}$ (red circle) are plotted an a function of the angle of magnetic field. The solid lines are merely guides to eyes.}
\end{center}
\end{figure}

To discuss more quantitative aspect, we plot as a function of $\theta$ the integrated intensities of the nonreciprocity   
\begin{eqnarray}
\Delta I_{12}=\int (S_{12}-S_{21}) df,
\end{eqnarray}
for the large $k$ modes of LiFe$_5$O$_8$ and Y$_3$Fe$_5$O$_{12}$ in Fig. 4 ($f$ is frequency). The integral intervals are 5 GHz $\leq f \leq$ 6 GHz for LiFe$_5$O$_8$ and 4.3 GHz $\leq f \leq$ 5.3 GHz for Y$_3$Fe$_5$O$_{12}$. The $\Delta I_{12}$ for Y$_3$Fe$_5$O$_{12}$ is negligible at $\theta =0$ and increases linearly with $\theta$. This behavior can be expected by the nonreciprocity induced by the magnetostatic surface mode (Damon-Eshbach mode)\cite{stancil}. The Damon-Eshbach mode is mediated by the magnetic dipole-dipole interaction on the surface of ferromagnetic sample, which is nonreciprocal mode propagating perpendicular to the magnetization. On the other hand, the nonreciprocity induced by the asymmetric band dispersion originating from Dzyaloshinskii-Moriya interaction is expected to shows the maximum around $\theta=0$ and decreases with increasing or decreasing $\theta$. The $\theta$ dependence of $\Delta I_{12}$ for LiFe$_5$O$_8$ seems to be explained by the combination of the Dzyaloshinskii-Moriya interaction and the Damon-Eshbach mode. Hence, the orign of large nonreciprocity at $\theta=0$ for LiFe$_5$O$_8$ can be ascribed to the Dzyaloshinskii-Moriya caused by the lack of inversion symmetry in this material.

In summary, we have observed the nonreciprocal magnon propagation due to the asymmetric band dispersion caused by the Dzyaloshinskii-Moriya interaction in a noncentrosymmetric ferromagnet LiFe$_5$O$_8$. The nonreciprocity seems useful for the manipulation of magnon spin current and microwave.

\section*{Methods}
Single crystals of LiFe$_5$O$_8$ were grown by the flux method\cite{Beregi1979}. 
Single crystals of Y$_3$Fe$_5$O$_8$ grown by Crystal Systems Corporation were used. 
The microwave response was measured in a superconducting magnet with use of a network analyzer. The detail of microwave experiment is described in the main text. All the experimental data in this paper was taken at room temperature.

\section*{Acknowledgements}
The authors thank Y. Nakamura, H. Toida, and S. Murakami for fruitful discussion.  This work was in part supported by the Grant-in-Aid for Scientific Research (Grant No. 25247058), Mitsubishi foundation, and Yamada Science Foundation.

\section*{Author contributions}
Y. I. and S. U. carried out the crystal growth, the photo-lithography, microwave measurement, and analyzed data. K. U. partly contributed to the photo-lithography. Y. O. planned and supervised the project. Y. I and Y. O. wrote the paper through the discussion and assistance from S. U. and K. U.

%\section*{Additional information} 
%The authors declare no competing financial interests.  Reprints and permissions information is available online at http://www.nature.com/reprints. 
%Correspondence and requests for materials should be addressed to Y. O. (c-onose@mail.ecc.u-tokyo.ac.jp)

\newpage
\section*{\large Supplementary information for the article \lq\lq Nonreciprocal magnon propagation in a noncentrosymmetric ferromagnet\rq\rq}
\subsection*{Magnetic property}
Figure 5 shows the magnetization curves of the LiFe$_5$O$_8$ sample, which is used in the microwave experiment, for various magnetic field directions. As shown in the inset, the sample shows rectangular shape. The longest direction is parallel to the [$1\bar{1}0$] direction, and the widest surface is perpendicular to the [111] direction. All the magnetization curves do not show hysteresis and saturate in a small magnetic field ($\sim 0.2$ T). These indicate that the magnetic anisotropy is small in this material. The small anisotropy of magnetization curves can be ascribed to the demagnetization effect.  

We show the microwave absorption spectra $\Delta S_{11}$ at various magnetic fields along [$1\bar{1}0$] for Y$_3$Fe$_5$O$_{12}$ and LiFe$_5$O$_8$ in Figs. 6 a,b. The $\Delta S_{11}$ spectra are deduced by the field-change of reflectance $S_{11}$ of sample-placed coplanar waveguide, similarly to the previous paper$^{ 1}$. For Y$_3$Fe$_5$O$_{12}$, $\Delta S_{11}$ shows sharp peak around 4.9 GHz and broad continuum above 5 GHz at 0.10 T. The frequencies increase with magnetic field but the spectral shape is almost unchanged. Similar spectra were already reported by An {\it et al}$^{ 2}$. The origin of sharp peak was ascribed to the uniform ($k=0$) ferromagnetic excitation (ferromagnetic resonance, FMR), and the continuum to the Damon-Eshbach mode. Similar spectral shape and magnetic field dependence are also observed for LiFe$_5$O$_8$ while the frequency scale is higher than that for Y$_3$Fe$_5$O$_{12}$. We plot in Fig. 6\textbf{c} the magnetic field dependence of FMR frequency for Y$_3$Fe$_5$O$_{12}$ and LiFe$_5$O$_8$. For both the samples, the FMR frequency increase linearly with the magnetic field. The frequency difference between Y$_3$Fe$_5$O$_{12}$ and LiFe$_5$O$_8$ may be caused by the differences in the strength of magnetic dipole-dipole interaction as well as magnetic anisotropy.

\subsection*{Analysis of microwave transmittance}

Figures 7a,b show the microwave transmittance $S_{21}$ from the port 1 to the port 2 in the microwave circuit shown in Fig. 1c. Here, the LiFe$_5$O$_8$ sample is placed at the center of the circuit. At 1.0 T, all the magnetic excitations are far above 10 GHz. Therefore, the $S_{21}$ is thought to be the background signal reflecting the details of microwave circuit besides the sample. When the magnetic field is lowered to 0.15 T or 0.20 T, the $S_{21}$ is increased because of the magnon propagation. In order to estimate the microwave transmittance owing to the magnon propagation, we deduce the difference of $S_{21}$ from the 1.0 T data, $\Delta S_{21}$. Figures 7\textbf{c,d} exemplify the deduced $\Delta S_{21}$.
Figure 8 illustrates the result of the same analysis for the Y$_3$Fe$_5$O$_{12}$ sample.

\begin{enumerate}
\item
Onose, Y., Okamura, Y., Seki, S., Ishiwata, S. \& Tokura, Y. Observation of Magnetic Excitations of Skyrmion Crystal in a Helimagnetic Insulator Cu$_2$OSeO$_3$. {\it Phys. Rev. Lett.} {\bf 109,} 037603 (2012).

\item
An, T. {\it et al.} Unidirectional spin-wave heat conveyer. {\it Nature Mater.} {\bf 12,} 549-553 (2013).
\end{enumerate}
\clearpage

\begin{figure}
\begin{center}
\includegraphics*[width=13cm]{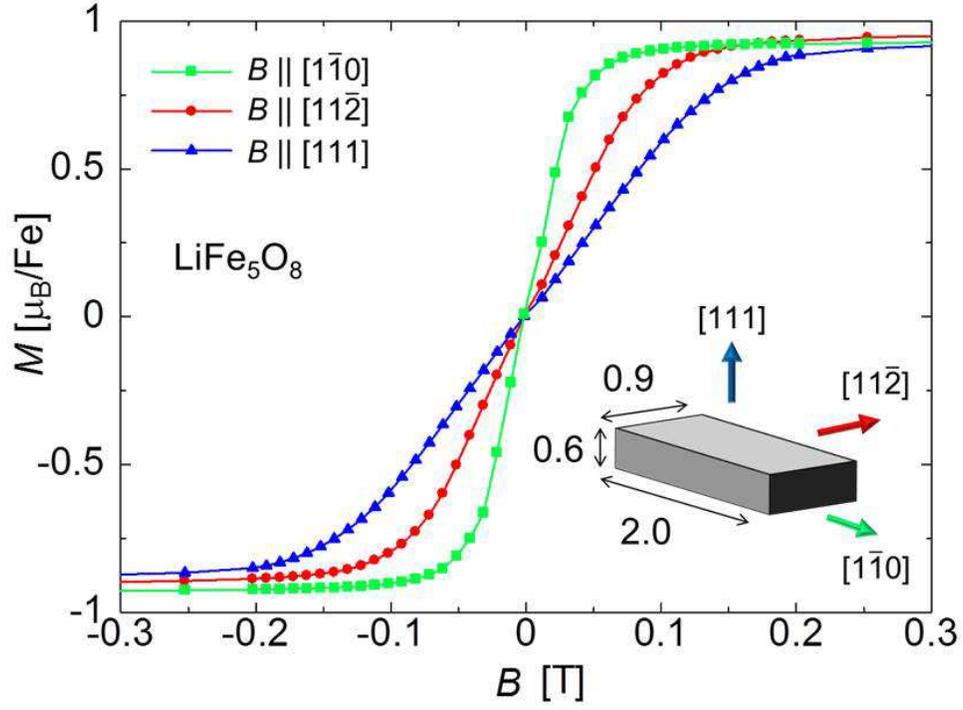}
\caption{Magnetization curves for the experimentally used LiFe$_5$O$_8$ sample for various magnetic field directions. Inset illustrates the sample shape and crystal orientation. }
\end{center}
\end{figure}

\begin{figure}
\begin{center}
\includegraphics*[width=15cm]{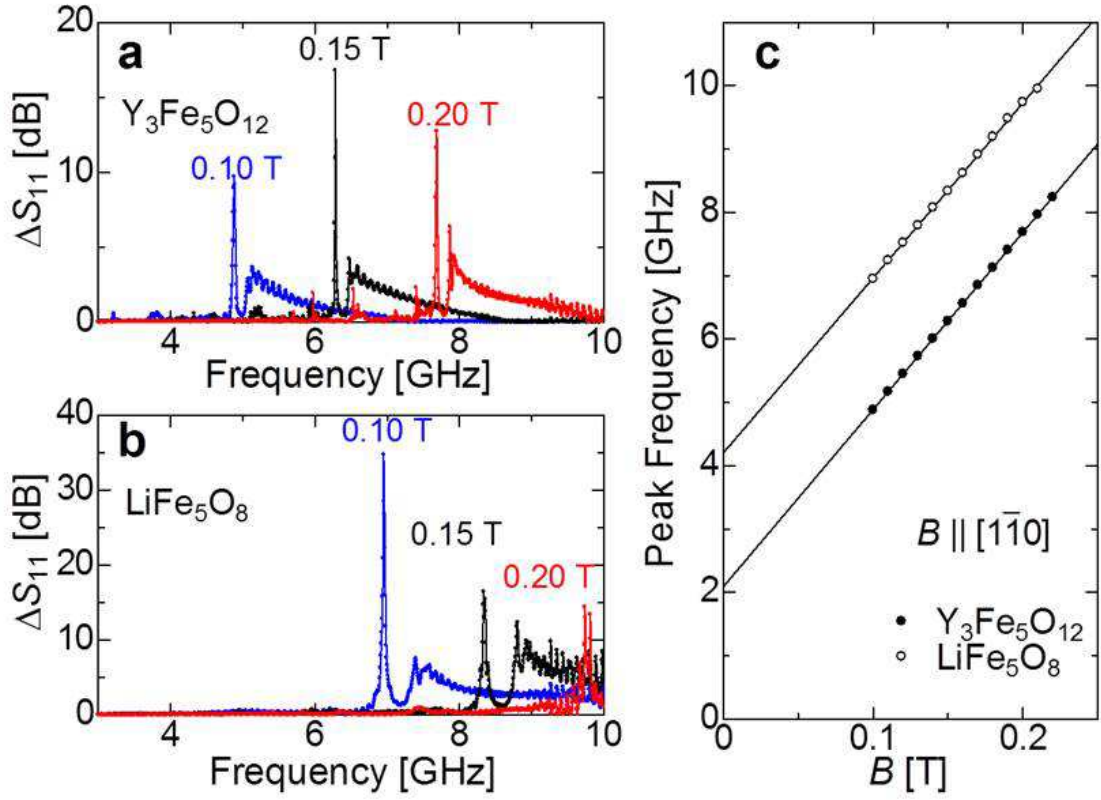}
\caption{\textbf{a,b} Microwave absorption spectra $\Delta S_{11}$ for (a) LiFe$_5$O$_8$ and (b) Y$_3$Fe$_5$O$_{12}$ in various magnetic fields along [$1\bar{1}0$]. \textbf{c} Magnetic field dependence of peak frequency for LiFe$_5$O$_8$ (open circle) and Y$_3$Fe$_5$O$_{12}$ (closed circle).}
\end{center}
\end{figure}

\begin{figure}
\begin{center}
\includegraphics*[width=15cm]{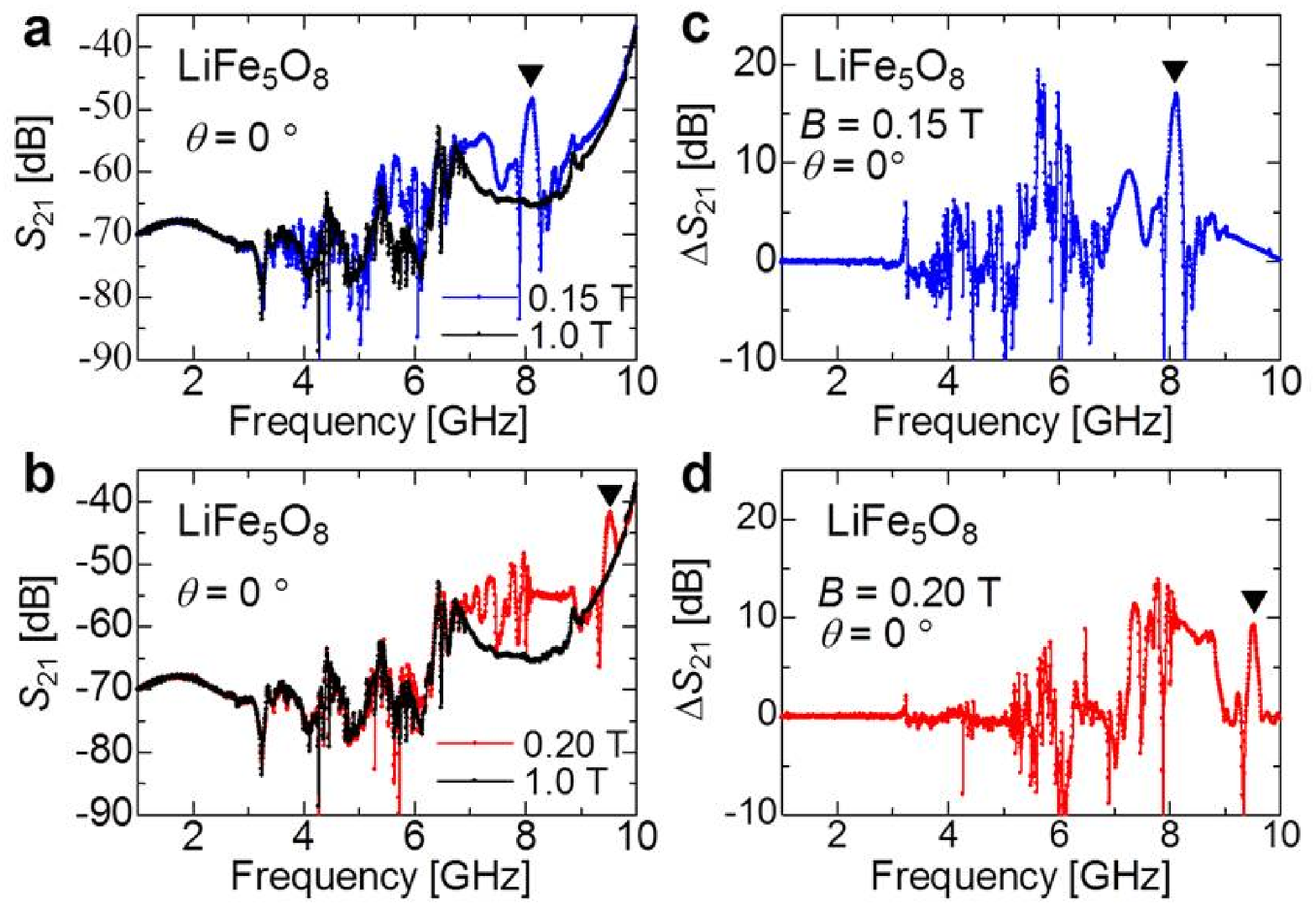}
\caption{\textbf{a,b} The microwave transmittance $S_{21}$ of the LiFe$_5$O$_8$ sample-placed microwave circuit at 0.15 T, 0.20 T, and 1.0 T ($\theta =0$). \textbf{c,d} The microwave transmittance owing to magnon propagation $\Delta S_{21}$ at (c) 0.15 T and (d) 0.20 T deduced from the difference of $S_{21}$ from the 1.0 T data. The inverted triangles show the FMR modes.}
\end{center}
\end{figure}

\begin{figure}
\begin{center}
\includegraphics*[width=15cm]{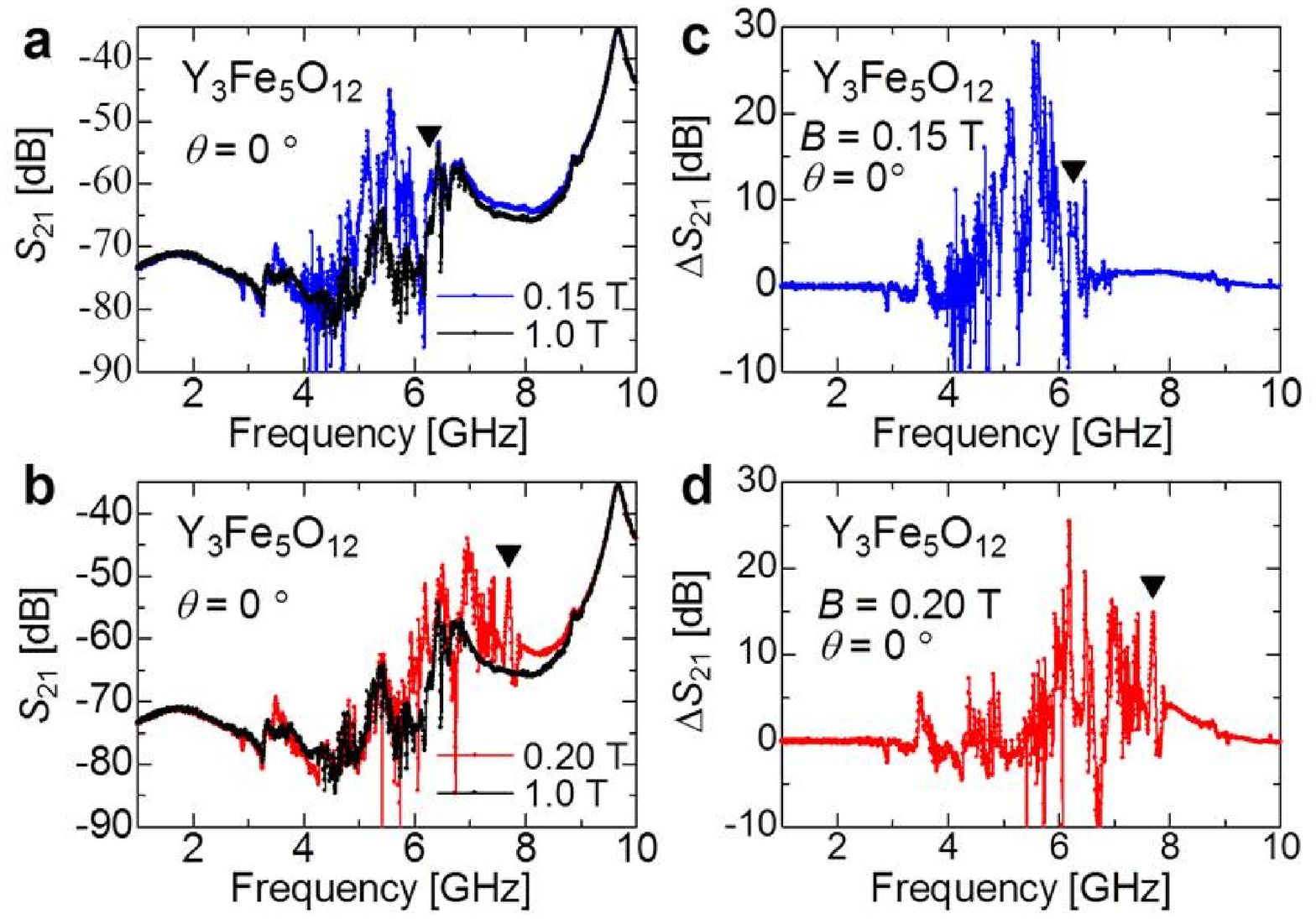}
\caption{\textbf{a,b} The microwave transmittance S$_{21}$ of the Y$_3$Fe$_5$O$_{12}$ sample-placed microwave circuit at 0.15 T, 0.20 T, and 1.0 T ($\theta =0$). \textbf{c,d} The microwave transmittance owing to magnon propagation $\Delta S_{21}$ at (c) 0.15 T and (d) 0.20 T deduced from the difference of $S_{21}$ from the 1.0 T data. The inverted triangles show the FMR modes.}
\end{center}
\end{figure}

\end{document}